\documentclass[11]{article}
\usepackage{geometry}
\usepackage{graphicx}
\usepackage{amsmath}
\usepackage{hyperref}
\usepackage{listings}
\usepackage{booktabs}
\usepackage[affil-it]{authblk}
\usepackage{orcidlink}

\title{OxO2 - A SSSOM Mapping Browser for Logically Sound Crosswalks}
\author[1,2]{Henriette Harmse \orcidlink{0000-0001-7251-9504}}
\author[1]{Haider Iqbal}
\author[1]{Helen Parkinson \orcidlink{0000-0003-3035-4195}}
\author[1]{James McLaughlin \orcidlink{0000-0002-8361-2795}}
\affil[1]{Samples, Phenotypes and Ontologies Team (SPOT), EMBL-EBI, Wellcome Genome Campus, Hinxton, CB10 1SD, Cambridgeshire, United Kingdom}
\affil[2]{Corresponding author: henriette007@ebi.ac.uk}

\date{September 2025}

\begin{document}
	\maketitle
	
	\begin{abstract}
EMBL-EBI created OxO to enable users to map between datasets that are annotated with different ontologies. Mappings identified by the first version of OxO were not necessarily logically sound, lacked important provenance information such as author and reviewer, and could timeout  or crash for certain requests. In this paper we introduce OxO2 to address these concerns. Provenance is addressed by implementing SSSOM, a mapping standard that defines provenance for mappings. SSSOM defines the conditions under which logical sound mappings can be derived and is implemented in OxO2 using Nemo, a Datalog rule engine. To ensure reasoning is performant and memory efficient, Nemo implements a number of strategies that ensures OxO2 will be stable for all requests. Due to these changes, OxO2 users will be able to integrate between disparate datasets with greater confidence.
	\end{abstract}
	
	\section*{Keywords}
	OxO; mapping; crosswalk; ontologies; FAIR Principles

\section{Introduction} \label{Introduction}
The FAIR principles are a set of guidelines designed to make datasets and their metadata Findable, Accessible, Interoperable and Reusable for both machines and humans~\cite{Wilkinson2016}, and require the use of FAIR vocabularies and/or ontologies \cite{Wilkinson2016, Xu2023}. Since their publication, the FAIR principles have been widely adopted to support FAIR data implementations~\cite{VanReisen2020}. 

Specific domains may aim to standardize on the ontologies they use, but often multiple ontologies are employed within a given domain~\cite{Jupp2017}. At EMBL-EBI, the Experimental Factor Ontology (EFO)~\cite{Malone2010} is used to annotate EMBL-EBI datasets. However, many other ontologies are also used within EMBL-EBI, such as the Mondo Disease Ontology (MONDO)~\cite{Vasilevsky2020}, the Human Disease Ontology (DOID)~\cite{Schriml2022} and the Human Phenotype Ontology (HP)~\cite{Kohler2021}. 

When considering cross-domain integration, differences in frames of reference mean that different domains often standardize on different ontologies~\cite{Vogt2025}. As an example, phenotypes (observable traits) are critical for diagnosing and treating diseases, and are studied in both model organisms (organisms that can easily be studied in a laboratory) such as zebrafish and mammals, as well as in humans. As such, zebrafish data are annotated with the Zebrafish Phenotype Ontology (ZP), mammalian data with the Mammalian Phenotype Ontology (MP) and human data with the Human Phenotype Ontology (HP), respectively. To study a specific phenotype, say increased heart size, researchers benefit from being able to combine data annotated with \texttt{ZP:0000532} (increased size of heart), \texttt{MP:0000274} (enlarged heart) and \texttt{HP:0001640} (Cardiomegaly)~\cite{Matentzoglu2025}. 

In order to facilitate these kinds of integration across datasets annotated with different ontologies, there is the need to map the concepts from one ontology to the related concepts of another ontology. EMBL-EBI created the Ontology Xref (Cross-reference) Service - referred to as OxO - to address this need \cite{Jupp2017}. In this paper we will refer to this version of OxO as OxO1.

OxO1 identified mappings between concepts of different ontologies by crawling the Ontology Lookup Service (OLS)~\cite{McLaughlin2025} and reading the \textbf{xref} (cross-reference relationships) annotations for each of the ontology concepts. The notion of an xref is intentionally loosely defined as part of the Open Biological and Biomedical Ontology (OBO) Flat File Format Specification, which has historically been used to define ontologies in the bioinformatics community~\cite{DayRichter2006,Laadhar2020,Smith2007}. An xref mapping may be used to indicate an exact match, but it could as well refer to a broader or narrower match. In fact xrefs may have legio meanings.

A key feature of OxO1 is that it enables walking across ontologies -- from a concept in one ontology to a concept in the next ontology -- referred to as a \textbf{crosswalk}. This is illustrated in Figure~1, where the crosswalk between \texttt{MONDO:0004981} and \texttt{DOID:1579} is realized by mapping from \texttt{MONDO:0004981} to \texttt{EFO:0000275}, to \texttt{OMIM:615770}, and finally to \texttt{DOID:1579}. This is referred to as a mapping at a \textbf{distance} of 3, since mappings between ontologies were performed 3 times. Distances greater than 3 are not supported in OxO1~\cite{Jupp2017}. Since xrefs are not well defined, the meaning of these kinds of crosswalks across multiple ontologies are also weakly defined.

Despite the loose definition of mappings in OxO1, it supports the following user groups.
\begin{description}
	\item \textbf{Researchers}, such as biologists seeking new treatments for heart disease, who may want to study the phenotype ``enlarged heart''.
	\item \textbf{Resource providers or data analysts} integrating datasets annotated with different ontologies, such as ZP and MP. 
\end{description}
At EMBL-EBI, OxO1 is used by resources such as European Variation Archive (EVA)~\cite{Cezard2022}, the Genome-Wide Association Study (GWAS) Catalog~\cite{Cerezo2025} and Europe PMC~\cite{EuropePMC2015}.  

\begin{figure}
	\centering
	\includegraphics[trim = 10mm 240mm 40mm 28mm, clip, scale=1]{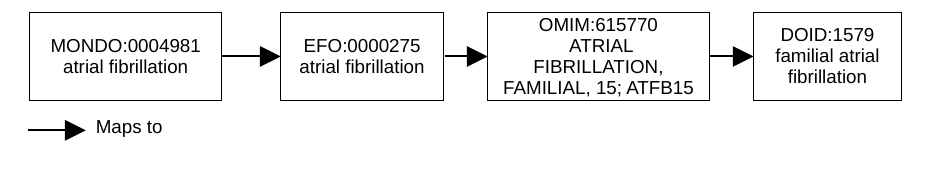}
	\caption{A crosswalk from \texttt{MONDO:0004981} to \texttt{DOID:1579} by mapping from \texttt{MONDO:0004981} to \texttt{EFO:0000275} to \texttt{OMIM:615770} to \texttt{DOID:1579}.  }
\end{figure}

The Simple Standard for Sharing Ontological Mappings (SSSOM) provides precise
definitions for mappings, along with related provenance information, and clearly defines the
conditions under which new mappings can be derived from existing mappings~\cite{Matentzoglu2022a}. In this paper we introduce OxO2, a new ontology mapping service that implements SSSOM and enables browsing of mappings between ontologies with logically sound crosswalks.

In the next section (Section~\ref{Background}), we give a brief overview of SSSOM and explain how it enables logically sound crosswalks. Section~\ref{Results} describes the value proposition of OxO2. In Section~\ref{Methods}, we motivate the design decisions of the OxO2 implementation, and in Section~\ref{Related}, we review related SSSOM implementations. Finally, in Section~\ref{Discussion}, we discuss the current limitations of our OxO2 implementation and future developments under consideration. 

\section{Background} \label{Background}
In this section, we explain how OxO2 uses Datalog (Section~\ref{Datalog}) to derive logically sound and complete mappings based on SSSOM chain rules (Section~\ref{Chain Rules}),  in a performant and memory efficient manner~(Section~\ref{Optimization}).

\subsection{An Overview of SSSOM}

\begin{figure} \label{figSSSOM}
	\centering
	\includegraphics[trim = 20mm 215mm 10mm 28mm, clip, scale=0.95]{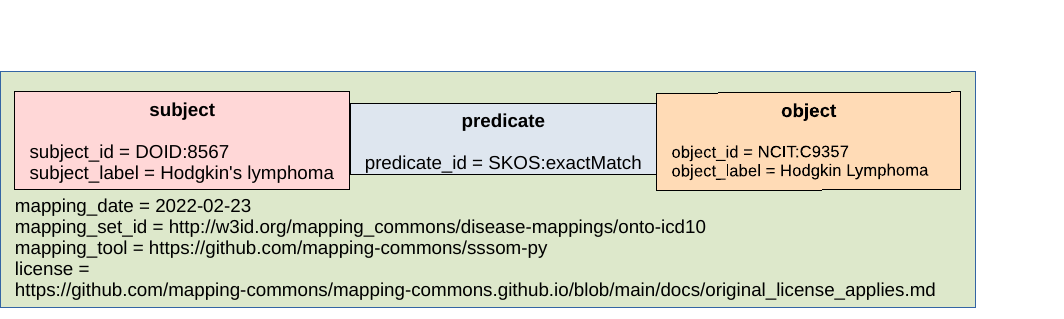}
	\caption{A SSSOM mapping that maps \texttt{DOID:8567} via \texttt{SKOS:exactMatch} to 
		\texttt{NCIT:C9357}. The pink, blue and orange boxes represent subject-, predicate- and object metadata. The green box represents the mapping set metadata. The pink, blue and orange boxes are inside the green box to indicate that mappings belong to a mapping set.}
\end{figure}

\begin{table*} 
	\caption{An abbreviated list of SSSOM standard metadata elements with example data, partially based on~\cite{DiseaseMappings}. Author and reviewer information are fictional.}
	\label{tab_SSSOM_metadata}
	\begin{tabular}{lp{5cm}p{7cm}} 
		\toprule
		Name & Description & Example\\
		\midrule
		author\_id & Identifies the persons or groups responsible for asserting the mappings & Henriette Harmse \\		
		confidence & A score between 0 and 1 denotes the confidence of correctness of a mapping. & 0.7\\
		mapping\_set\_id & A globally unique identifier for the mapping set & http://w3id.org/mapping\_commons/disease-mappings/onto-icd10 \\
		mapping\_tool & A reference to the tool or algorithm that was used to generate the mapping & https://github.com/mapping-commons/sssom-py \\
		object\_id & The ID of the object & \texttt{NCIT:C9357} \\
		object\_label & The label of the object & Hodgkin Lymphoma \\
		predicate\_id & The ID that relates subject and object & \texttt{SKOS:exactMatch} \\
		reviewer\_id & Identifies the persons or groups that reviewed and confirmed the mapping &  Helen Parkinson \\		
		subject\_id & The ID of the subject of the mapping & \texttt{DOID:8567} \\
		subject\_label & The label of subject of the mapping & Hodgkin's lymphoma \\ 		
		\bottomrule
	\end{tabular}
\end{table*}

SSSOM~\cite{Matentzoglu2022a,Matentzoglu2022b,Matentzoglu2023,SSSOM} is briefly summarised here, along with an example to illustrate the value proposition of OxO2.

Mappings are defined as a subject mapping via a predicate to an object. In SSSOM, this is expressed as a \texttt{subject\_id} mapping via a \texttt{predicate\_id} to an \texttt{object\_id}, where
the \texttt{subject\_id}, \texttt{predicate\_id} and \texttt{object\_id} are each expressed  as abbreviated Uniform Resource Identifiers (URIs), called CURIEs~\cite{Birbeck2010}.
In general all identifiers in SSSOM are CURIEs, where CURIEs can refer to entities including ontology concepts, database references etc. To define the provenance of a mapping, a mapping is described using SSSOM standard metadata elements, such as \texttt{subject\_label}, \texttt{author\_id} (see Table~\ref{tab_SSSOM_metadata}).

\begin{table*} 
	\caption{An example SSSOM mapping file. }
	\label{tab_SSSOM_example}
	\begin{tabular}{llp{3.1cm}llp{3cm}}
		\toprule
		subject\_id	& subject\_label& predicate\_id	&	object\_id		&	object\_label\\
		\midrule
		DOID:8567 & Hodgkin's lymphoma & SKOS:exactMatch &	NCIT:C9357 & Hodgkin Lymphoma \\
		HP:0012189 & Hodgkin lymphoma	& OWL:equivalentClass	& DOID:8567 & Hodgkin's lymphoma\\
		MONDO:0009348	& classic Hodgkin lymphoma &	SKOS:closeMatch		& HP:0012189 &	Hodgkin lymphoma \\
		\bottomrule
	\end{tabular}
\end{table*} 


For ease of use, SSSOM defines a tab-separated values (TSV) table format for representing mappings. Table~\ref{tab_SSSOM_example} defines three mappings, firstly that \texttt{DOID:8567} maps via \texttt{SKOS:exactMatch} to \texttt{NCIT:C9357}, secondly that \texttt{HP:0012189} maps via \texttt{OWL:equivalentClass} to \texttt{DOID:8567}, and lastly that \texttt{MONDO:0009348} maps via \texttt{SKOS:closeMatch} to \texttt{HP:0012189}. To define the metadata of a set of mappings, a mappings TSV file can be prepended with a header in YAML (a human-friendly data serialization) format~\cite{YAML}. The corresponding header for Table~\ref{tab_SSSOM_metadata} is provided in Listing~\ref{lst:header}.

\begin{lstlisting}[label=lst:header,caption=YAML Header that defines the mapping set for the mappings in Table~\ref{tab_SSSOM_example}]
	# mapping_set_id: https://fois2025.com/example.mappings.tsv
	# curie_map:
	#   DOID: http://purl.obolibrary.org/obo/DOID_
	#   HP: http://purl.obolibrary.org/obo/HP_
	#   MONDO: http://purl.obolibrary.org/obo/MONDO_
	#   NCIT: http://purl.obolibrary.org/obo/NCIT_
	#   OWL: http://www.w3.org/2002/07/owl#
	#   SKOS: http://www.w3.org/2004/02/skos/core#
\end{lstlisting}

\subsection{Chain Rules} \label{Chain Rules}
With SSSOM it is possible to derive mappings from existing ones in a way that ensures derived mappings are logically sound. This is achieved by providing chain rules that define the circumstances under which mappings can be composed from existing mappings.  A \textbf{rule} defines the conditions under which an action will occur and follows an if-then structure. \textbf{Rule chaining} (or rule chains) refers to the mechanism where the execution of one rule triggers the execution of subsequent rules. Execution stops when no more rules are applicable~\cite{Abiteboul1995}. SSSOM chain rules are categorized as transitivity rules, rule chains over exact and equivalent matches, inverse rules and generalization rules. The approach that SSSOM takes is to provide reasonable defaults for mapping tools. Hence, SSSOM defines transitive rules for \texttt{skos:narrowMatch} and \texttt{skos:broadMatch}~\cite{SSSOMChain}, despite them not being defined as transitive in the Simple Knowledge Organization System (SKOS) specification~\cite{SKOS}. In total, the SSSOM chain rules define 22 rules that can be applied to SSSOM mappings, all of which are implemented and used in OxO2. 

For illustration purposes, we will focus only on chain rules over exact and equivalent matches, labeled as RCE2 in the SSSOM specification~\cite{SSSOMChain}. RCE2 has two variants, which apply to mappings where the \texttt{predicate\_id} is either \texttt{SKOS:exactMatch} or \texttt{OWL:equivalentClass}. These two variants are defined in Listings \ref{RCE2_1} and \ref{RCE2_2}.

\newpage
\begin{lstlisting}[label=RCE2_1,caption=RCE2-1, mathescape=true]
	$A-[p]-B$,
	$B-[\texttt{OWL:equivalentClass}]-C $
	$\rightarrow A-[p]-C$
\end{lstlisting}

\begin{lstlisting}[label=RCE2_2,caption=RCE2-2, mathescape=true]
	$A-[p]-B$,
	$B-[\texttt{SKOS:exactMatch}]-C $
	$\rightarrow A-[p]-C$
\end{lstlisting}

Listing \ref{RCE2_1} states that when we have CURIEs $A$, $B$ and $C$, such that $A$ is mapped to $B$ via an arbitrary \texttt{predicate\_id} $p$, and $B$ is mapped to $C$ via the \texttt{predicate\_id} \texttt{OWL:equivalentClass}, then we can derive that $A$ maps to $C$ via the same arbitrary \texttt{predicate\_id} $p$. The meaning of Listing~\ref{RCE2_2} follows in a similar fashion.

Based on the mappings in Table~\ref{tab_SSSOM_example}, we can derive the two new mappings shown in Listings~\ref{lst:derive1} and \ref{lst:derive2}, by applying the chain rules of Listing~\ref{RCE2_1} and~\ref{RCE2_2}, respectively. In Listing~\ref{lst:derive1} we derive that \texttt{MONDO:0009348} (classic Hodgkin lymphoma) has a \texttt{SKOS:closeMatch} with 
\texttt{DOID:8567} (Hodgkin's lymphoma), by applying the chain rule of Listing~\ref{RCE2_1}. By applying the chain rule of Listing~\ref{RCE2_2}, Listing~\ref{lst:derive2} derives that \texttt{MONDO:0009348} has a \texttt{SKOS:closeMatch} with \texttt{NCIT:C9357} (Hodgkin Lymphoma), by using both an existing mapping from Table~\ref{tab_SSSOM_example} and the derived mapping from Listing~\ref{lst:derive1}.

\begin{lstlisting}[label=lst:derive1,caption=Apply RCE2-1, mathescape=true]
	$\texttt{MONDO:0009348}-[\texttt{SKOS:closeMatch}]-\texttt{HP:0012189}, $
	$\texttt{HP:0012189}-[\texttt{OWL:equivalentClass}]-\texttt{DOID:8567}  $
	$\rightarrow \texttt{MONDO:0009348}-[\texttt{SKOS:closeMatch}]-\texttt{DOID:8567}$
\end{lstlisting}

\begin{lstlisting}[label=lst:derive2,caption=Apply RCE2-2, mathescape=true]
	$\texttt{MONDO:0009348}-[\texttt{SKOS:closeMatch}]-\texttt{DOID:8567},$ 
	$\texttt{DOID:8567}-[\texttt{SKOS:exactMatch}]-\texttt{NCIT:C9357}  $
	$\rightarrow \texttt{MONDO:0009348}-[\texttt{SKOS:closeMatch}]-\texttt{NCIT:C9357}$
\end{lstlisting}

\subsection{Datalog} \label{Datalog}
In this section we show that SSSOM chain rules can be expressed as Datalog rules, and hence, all characteristics of Datalog rules also apply to SSSOM chain rules. OxO2 uses Datalog to derive logically sound and complete mappings in a performant and memory efficient way.

Datalog is a declarative logic programming language designed for logical inference. A \textbf{Datalog rule} defines the premises from which a conclusion follows and has an if-then structure similar to chain rules. A \textbf{Datalog program} consists of Datalog rules. Datalog rules are applied recursively which means that the result of a rule can be used by other rules to derive new information~\cite{Abiteboul1995}. A Datalog rules engine is an application that, given a set of facts and a set of rules, it can infer new facts in a way that is logically sound. Since Datalog is declarative, it means that one can state the rules that are applicable, without having to state how the inferences are to be determined. The Datalog rules engine is responsible for implementing the method(s) of determining inferences~\cite{Ivliev2023}. 

To ensure that all facts that can be derived from a Datalog program are finite, 2 conditions must hold: firstly, all facts must be constants, and secondly, all variables in the conclusion of a rule, must appear in the body of a rule. We note that these conditions hold for SSSOM chain rules~\cite{Ceri1989}:
\begin{enumerate}
	\item All SSSOM mappings are without variables, as seen from the first mapping in Table~\ref{tab_SSSOM_example}, which states that \texttt{DOID:8567} maps via \texttt{SKOS:exactMatch} to \texttt{NCIT:C9357} where \texttt{DOID:8567}, \texttt{SKOS:exactMatch} and \texttt{NCIT:C9357} are all constants.
	\item The second condition holds for the chain rule defined in Listing~\ref{RCE2_1}. The conclusion, given after the $\rightarrow$, is $A-[p]-C$ where $A$, $p$ and $C$ are variables. These variables appear in the premises $A-[p]-B$ and $B-[\texttt{OWL:equivalentClass}]-C$ where $A$ and $p$ appear in the first premise and $C$ in the second premise. Doing a similar check, shows that condition 2 holds for Listing~\ref{RCE2_2} as well. Note that this condition only applies to variables and not constants. Hence, any constant, such as \texttt{SKOS:exactMatch} can appear in the conclusion without appearing in any premises. We note that for all SSSOM chain rules~\cite{SSSOMChain} this condition holds.
\end{enumerate} 

Datalog has some characteristics~\cite{Ceri1989} that are important for OxO2.
\begin{description}
	\item[Termination] For a finite set of facts, Datalog inferencing always terminates. Hence, the OxO2 dataloader can assume inferencing is complete when the inferencing algorithm terminates.
	\item[Soundness] All Datalog inferences are guaranteed to follow logically from the assumed facts. For OxO2 this means inferred mappings can be trusted to be logically correct.
	\item[Completeness] means that all inferences that can be derived from the facts, are indeed derived. Hence, OxO2 is guaranteed to include all mappings that follow logically from asserted mappings.
	\item[Favourable time and space computational complexity] The output of the Datalog inference algorithm, and the time it takes to run till it terminates, is polynomial in the size of the input. This means the time and space needs of Datalog inferencing is limited by a polynomial, such as $n^k$, where $n$ represents the input size and $k$ is a constant. This is a positive result when compared to time and space needs that can be exponential, e.g., $k^n$. To illustrate the difference between polynomial time and exponential time, assume the input size is $n=10$ and $k=2$, then the polynomial time is $n^k = 10^2 = 100$, while exponential time is $k^n=2^{10}=1024$. If we now change the input size to $n=100$, we see that the polynomial time is still reasonable, whereas the exponential time is not~\footnote{There are many other computational complexity classes besides polynomial and exponential. See~\cite{Davis1994} for details.}. This result indicates that theoretically the Datalog inference algorithm is feasible.
\end{description}

\subsection{Strategies for optimizing Datalog Reasoning} \label{Optimization}
Datalog reasoning has some challenges when applied to large knowledge graphs and hence a number of strategies are employed to deal with these challenges. We mention the strategies here that are used in the implementation of OxO2.

The most basic approach to Datalog inferencing is the Naive Evaluation algorithm. It starts with the original facts, and in the first step it derives new facts by applying all the rules to the original facts. Subsequent steps apply the rules to both original facts and derived facts, repeatedly, until no new facts can be added. This can result in the same inferences being made repeatedly. To avoid this redundancy, an optimization is to  apply subsequent rules only to facts derived in the previous iteration, which is called Seminaive Evaluation~\cite{Abiteboul1995}.

A related concern is the explanation of inferences,  which is known to be exponential in the number of facts. This is because there can be multiple explanations for a single inference. To avoid this explosion of the number of explanations, a technique called tracing is used. Instead of recording all possible explanations, tracing aims to provide only 1 explanation for a given inference. This approach ensures that the explanation of all inferences is feasible~\cite{Elhalawati2022}. 

To ensure fast and compact in-memory data storage, a column-based storage layout is used which enables efficient compression schemes. However, column-based storage increases the cost of updates (e.g. when new inferences are made). For this reason each rule application is written to a separate delta table~\cite{Urbani2016}. To reduce the cost of combining delta tables, combined deltas are cached~\cite{Ivliev2024}.  

Even with the above optimizations, users can wait a long time when inferences are done in realtime. Therefore, it is recommended to precompute inferences to improve the user experience~\cite{Urbani2016}.

\section{Results} \label{Results} 

\subsection{OxO1 Challenges} \label{Challenges}
The key value proposition of OxO1 is that it is able to identify potentially related mappings that could either map directly or indirectly up to a distance of 3 to other CURIEs. OxO1 has 3 main challenges.
\begin{enumerate}
	\item The semantics of the mappings it returns are vague, which is exacerbated for mappings at a distance greater than 1. In OxO1 it is possible to derive mappings that have the form of Listing \ref{lst:problematic}:   
	\begin{lstlisting}[label=lst:problematic,caption=Problematic inferred mapping of OxO1, mathescape=true]
		$A-[\texttt{SKOS:broadMatch}]-B$, 
		$B-[\texttt{SKOS:narrowMatch}]-C$  
		$ \rightarrow A-[\textbf{\texttt{meaning?}}]-C$
	\end{lstlisting}	
	\item Returning mappings at a distance of 3 is only feasible when requesting mappings
	for a small set of CURIEs. This problem became more accentuated as the number of
	mappings and ontologies in OxO1 grew, with queries either timing
	out, or the OxO1 server crashing. To find mappings for a CURIE at distance 1, OxO1 queries all mappings in 
	its Neo4J database. As these mappings are stored in the database, returning these results are highly efficient. For retrieving mappings at a distance of 2 or 3, OxO1 determines derived mapping at query time using the following algorithm. It first determines the cross product of all terms in the database. 
	This result is used to define possible starting and ending nodes, for which all possible paths of length 2 or 3 are determined, depending on the distance specified.
	The computational complexity of the cross product is $n \times n$, where $n$ is the number of terms. The computational complexity for determining all paths depends on the length of the path (or distance) $d$, and the average degree of the nodes of the graph under consideration, which is denoted by $k$. The degree of a node is determined by the number of inbound and outbound edges connected to it. The computational complexity for determining all paths is given by $k^d$. Hence, the total computational complexity for the OxO1 algorithm is $k^d \times n^2$. At the time of writing these values for OxO1 are as follows:
	\begin{itemize}
		\item the number of terms $n = 698~651$,
		\item the average degree of nodes $k = 3$, and
		\item the length of the paths to determine, which is also the mapping distance in OxO, where $d=2$ or $d=3$.
	\end{itemize}
	Hence, the worst case complexity at distance $2$ is $3^2 \times 698~650^2 = 4.393006402 \times 10^{12}$ and for distance $3$ it is $3^3 \times 698~650^2 = 1.317901921 \times 10^{13}$. Looking at these calculations it becomes clear why OxO1 crashed or timeout for large requests.
\end{enumerate}

\subsection{How OxO2 address these challenges}

To address the concerns regarding semantics, OxO2 implements SSSOM fully, and hence all metadata associated with mappings and mapping sets, are stored in OxO2. Its Application Programming Interface (API) allows searching across all SSSOM metadata elements (see Table~\ref{tab_SSSOM_metadata}). This provides data integrators with multiple ways in which they can find mappings, thereby increasing their chances to find the data most suitable to their needs.

In order to ensure inferences are logically sound, OxO2 makes use of a Datalog rule engine to infer mappings. Hence, OxO2 users can trust that inferred mappings are logically correct. To enable fast searches across inferred mappings, OxO2 materializes all chain rule inferences as part of its data release (see Section~\ref{Optimization}). Moreover, there is no longer a restriction on the distance at which mappings can be returned, or the length of crosswalks that can be identified by OxO2. The derived mappings of Listings~\ref{lst:derive1} and~\ref{lst:derive2} were derived using this approach, and the mapping sets from \cite{DiseaseMappings}. Mappings that are derived by OxO2, are placed in a mapping set with value
\texttt{https://www.ebi.ac.uk/spot/oxo/inferences}, \texttt{mapping\_justification} value
\texttt{SEMAPV:MappingChaining} and \texttt{mapping\_tool} value as \texttt{OxO2}.

SSSOM's 22 chain rules are helpful to define the precise conditions under which new mappings can be inferred that are logically sound. But because chain rules can be recursively applied multiple times, how inferred mappings were derived may not be obvious to humans. To make this information accessible to users, one needs to identify all the chain rules that have been applied, and the facts that were used, to derive an inferred mapping. We are able to provide this information by using a Datalog rule engine, that records this information as part of OxO2's data release process. 

\subsection{Benefits to Users}

OxO2 will help our users in the following ways:
\begin{enumerate}
	\item A researcher searching for "enlarged heart" will be able to find the MP concept \texttt{MP:0000274}, but they may be missing \texttt{ZP:0000532} (increased
	size of heart) and \texttt{HP:0001640} (Cardiomegaly). OxO2 will help them to uncover \texttt{ZP:0000532} and \texttt{HP:0001640}.
	
	\item Logical soundness does not necessarily mean biological correctness. Assume for the moment a mapping was mistakenly added that states that 
	\texttt{MP:0000274} (enlarged heart) is a \texttt{SKOS:exactMatch} with \texttt{MP:0001095} (enlarged trigeminal ganglion). This could result in inferred mappings that are logically sound, but which are nonsensical from a biological perspective. A researcher can identify an error like this by verifying the premises of an inference. As an example, we can verify the inferences of Listings~\ref{lst:derive1} and~\ref{lst:derive2} by verifying the premises in Table~\ref{tab_SSSOM_example}. 
	
	\item Derived mappings in OxO1 were not axiomatised, and hence, the only way researchers could verify each of the mappings, was to manually validate them. In the case of OxO2, one can assume that the inferred mappings are logically correct. Any incorrect inferred mappings will only ever be due to incorrect explicitly stated mappings. Thus, to validate the correctness of derived mappings, a user can consider reviewing explicitly stated mappings. 
	
	\item Complete explanations of inferred mappings will help data integrators to identify the source of incorrect data, since explanations clearly state the facts that were used to derive inferred mappings. 
	
\end{enumerate}

\section{Methods} \label{Methods}

For materializing inferred mappings, OxO2 uses Nemo, an in-memory rule engine \cite{Ivliev2023,Ivliev2024}. Nemo runs as a command line tool that loads facts and rules, 
processes them, and, once complete, writes out the relevant inferences; then the Nemo process terminates. From this perspective Nemo is well aligned with being run as part of a data release pipeline, as is the case for OxO2. To ensure that inferencing is fast and memory efficient, Nemo implements the various optimizations mentioned in Section~\ref{Optimization}. 

Nemo implements Datalog with various extensions, that results in inferencing not necessarily terminating (Section~\ref{Datalog}). However, OxO2's use of Nemo is such that it only uses Datalog without any extensions and, hence for its use case, Nemo terminates~\cite{Ivliev2024}.

Currently OxO2 imports $1~160~020$ mappings, from which it is able to infer
$49~536$ mappings. Inferencing on an Intel Core Ultra 7 165U x 13 laptop, with 32GB
RAM and SSD, completes in about 17 min, and uses about 380MB of memory. These results show that it is possible to run an OxO2 dataload on a personal computer, without necessarily having access to high-performance computing infrastructure.

For storing mappings, OxO2 uses Solr rather than Neo4J. Our motivation for this change in architecture is due to there being no need for graph queries in OxO2 -- all derived mappings with their explanations are determined during the OxO2 dataload. All inferred mappings and their explanations are then stored in Solr. The OxO2 Solr schema design mimics the SSSOM metadata elements with a core for storing mapping information and core for storing mapping set information. Hence, all SSSOM metadata elements are searchable and can be accessed via API calls and its frontend. Explanations are stored along with the mappings in the mapping core. 

The OxO2 dataload and backend are implemented using Java 17 and Spring Boot. The frontend is implemented in React using Typescript. Styling is implemented using TailWind CSS, which provides numerous utility classes that limits the need for custom CSS. 

\section{Related Tools} \label{Related}
In support of SSSOM there are a number of tools as detailed in \cite{Matentzoglu2023}:
\begin{itemize}
	\item \textbf{sssom-py} is a Python library for reading, transforming and manipulating SSSOM mappings.
	\item The \textbf{Ontology Access Kit (OAK)} can provide mappings in the SSSOM format.
	\item The \textbf{sssom-java library} is a Java library for reading, transforming and manipulating SSSOM mappings.
	\item \textbf{Semantic Reasoning Assembler (SeMRA)} is a Python library that allows one to crawl mappings in various formats, not just SSSOM. It then de-duplicates mappings, infers mappings, derives associated confidence for inferred mappings, and filters false positive mappings \cite{Hoyt2025}. 		
\end{itemize} 

These tools are used for the creation and manipulation of SSSOM files. The main purpose of
OxO2 is to make existing mappings and their derived mappings available for users to search
and browse.

The value of SSSOM as a standard for mappings depends on its community adoption. Currently most mappings in the bioinformatics domain are still embedded within ontologies, which are extracted by the OLS dataload into SSSOM files that are imported by OxO2. But many of these mappings are based on xrefs and hence still suffer from the poor semantics we discussed in the Introduction. For this reason, the tools listed above are complementary to OxO2, in that they enable users to create high quality mappings that are essential to the success of OxO2.

\section{Discussion and Future} \label{Discussion}
With the development of OxO2, we are able to address the key limitations of OxO1 in terms of the semantics of mappings, the meaning of inferred mappings and the length of crosswalks it can support. It is only through the availability of SSSOM that OxO2 is able to address these concerns.

Currently OxO2 is still in active development and we will prioritise as follows:
\begin{enumerate}
	\item Currently, the API of OxO2 focuses on bringing new search capabilities to users, with
	its own new request and response structures. For our existing users there may be a
	need to create a backwards compatibility layer, so as to not disrupt their services
	when OxO2 is rolled out. Similar to OLS4~\cite{McLaughlin2025}, this will be
	implemented using view classes that transforms the underlying OxO2 data model into the OxO1 data model view.
	\item The value proposition of OxO2 is dependent on high quality SSSOM mappings. We need to identify mappings sets that can be included in OxO2 such as mappings from MONDO to HP found at~\cite{MONDO2HP} and SeMRA which provides an SSSOM export of their mappings~\cite{SeMRAMappings}. 
	\item The explanation implementation is currently a proof of concept and contains only information on the \texttt{subject\_id}, \texttt{predicate\_id}, \texttt{object\_id} and chain rule that has been applied. To provide sufficient provenance, it also needs to include information such as \texttt{author\_id}, \texttt{reviewer\_id}, and the mapping set from which premises came. Currently, explanations are given in a text form. This needs to be 	augmented with a graph that can easily be navigated by users.
	\item To ensure the high availability of OxO2, its data release pipeline will be rolled out on the EMBL-EBI high-performance computing infrastructure, and its frontend and backend to the Kubernetes infrastructure. 
	\item The EMBL-EBI instance of OxO is focussed on the biological domain. To enable other domains to install their own instances of OxO2, OxO2 will be dockerized.
\end{enumerate}
	

\section*{Online Resources}
\textbf{OxO2 code base: }\href{OxO2 code base}{https://github.com/EBISPOT/oxo2}

\section*{Acknowledgements}
J.A.M., H.I., H.P., and H.H. are supported in part by EMBL-
EBI Core Funds. J.M., H.I. and H.H are supported in part by EVORA. The EVORA project has received funding from the European Union's HORIZON programme under grant agreement No 101131959. 

For the purpose of Open Access, a CC-BY public copyright licence has been applied to the present document and will be applied to all subsequent versions up to the Author Accepted Manuscript arising from this submission.

\end{document}